\documentclass[aps,pre,twocolumn,showpacs,superscriptaddress,preprintnumbers,amsmath,amssymb]{revtex4}
\usepackage{amsmath}
\usepackage{graphicx}
\usepackage{dcolumn}
\usepackage{bm}
\usepackage{float}

\begin{document}

\title{A thin-shell instability in collisionless plasma}

\author{M. E. Dieckmann}
\affiliation{Department of Science and Technology, Link\"oping University, SE-60174 Norrk\"oping, Sweden}

\author{H. Ahmed}\affiliation{Centre for Plasma Physics (CPP), Queen's University Belfast, BT7 1NN, UK}
\author{D. Doria}\affiliation{Centre for Plasma Physics (CPP), Queen's University Belfast, BT7 1NN, UK}
\author{G. Sarri}\affiliation{Centre for Plasma Physics (CPP), Queen's University Belfast, BT7 1NN, UK}
\author{R. Walder}\affiliation{\'Ecole Normale Sup\'erieure, Lyon, CRAL, UMR CNRS 5574, Universit\'e de Lyon, France}
\author{D. Folini}\affiliation{\'Ecole Normale Sup\'erieure, Lyon, CRAL, UMR CNRS 5574, Universit\'e de Lyon, France}
\author{A. Bret}\affiliation{ETSI Industriales, Universidad Castilla La Mancha, 13 071 Ciudad Real, Spain}
\author{A. Ynnerman}
\affiliation{Department of Science and Technology, Link\"oping University, SE-60174 Norrk\"oping, Sweden}
\author{M. Borghesi}\affiliation{Centre for Plasma Physics (CPP), Queen's University Belfast, BT7 1NN, UK}

\date{\today}


\begin{abstract}
The thin-shell instability has been named as one process, which can generate entangled structures in astrophysical plasma on collisional (fluid) scales. It is driven by a spatially varying imbalance between the ram pressure of the inflowing upstream plasma and the downstream's thermal pressure at a non-planar shock. Here we show by means of a particle-in-cell (PIC) simulation that an analogue process can destabilize a thin shell formed by two interpenetrating, unmagnetized and collisionless plasma clouds. The amplitude of the shell's spatial modulation grows and saturates after about ten inverse proton plasma frequencies, when the shell consists of connected piecewise linear patches. 
\end{abstract}

\maketitle
A shock wave forms if an obstacle moves through a medium at a speed, which exceeds the speed at which the medium can convey information. Shocks are ubiquitous in fluids and plasmas. The bow shock that separates the solar wind from the Earth's magnetopause is probably the best-understood plasma shock \cite{Bale05,Burgess05}. Even larger shocks can form where energetic plasma outflows collide with the interstellar medium. The solar wind termination shock separates the heliosphere from the interstellar medium \cite{Richardson08}. The existence of plasma shocks outside of the solar system, such as supernova remnant shocks \cite{Truelove99}, can be inferred from the electromagnetic radiation they emanate. A comprehensive review of these non-relativistic magnetized shocks is given by Ref. \cite{Treumann09}. 

The properties of shocks are determined by those of the underlying medium; the most important one being the particle's mean free path. If the mean-free path of the particles is small compared to the shock scale, then we speak of a fluid- or hydrodynamic shock.

A fluid shock is a discontinuity that separates a pre-shock (upstream) fluid from a post-shock (downstream) fluid. The velocity component of the upstream flow along the shock normal and measured in the reference frame of the shock is supersonic. This velocity component is reduced to a subsonic value by the shock crossing and the released directed flow energy is transformed into internal energy. The velocity component of the fluid, which is orthogonal to the shock normal, is left unchanged by the shock crossing. If the shock is planar and the lateral velocity constant, then we can remove the lateral velocity by choosing a comoving reference frame.

Let us assume that the upstream velocity is spatially uniform and that the shock front has a sinusoidal modulation. The direction of the shock normal and, hence, the components of the upstream velocity vector along this normal and orthogonal to it vary along the shock boundary. We can not find a reference frame, in which the lateral velocity vanishes everywhere. The normal component of the fluid velocity is reduced by the shock crossing while the lateral velocity remains unchanged, which implies a rotation of the fluid velocity vector. The rotation angle varies with the position along the shock boundary. This spatially varying rotation of the velocity vector implies a spatially varying momentum transfer from the upstream fluid to the shock boundary and, hence, a spatially varying imbalance between the ram pressure of the upstream flow and the thermal pressure of the downstream fluid. Such an imbalance can yield a linear \cite{Vishniac83} or a nonlinear \cite{Vishniac94} instability in the case of a thin fluid shell that is enclosed by two shocks. The thin-shell instability (TSI) amplifies the corrugation of the shock boundaries. The nonlinear TSI discussed in Ref. \cite{Vishniac94} has been examined by means of hydrodynamic \cite{Blondin96} and magnetohydrodynamic \cite{Heitsch07} simulations. 

Hydrodynamic \cite{Eidmann00,Le00} and particle-in-cell (PIC) \cite{Silva04,dHumieres05,He07} simulations demonstrate that the ablation of a target by an intense laser pulse generates an energetic plasma flow. The collision of this plasma flow with ionized residual gas, which is usually collisionless, gives rise to the electrostatic shocks that have been detected in many simulations and experiments \cite{Dean71,Forslund70,Forslund71,Biskamp73,Gitomer86,Bell88,Romagnani08,Kuramitsu11,Ahmed13}. The laser-plasma experiment performed in Ref. \cite{Ahmed13} resulted in a thin plasma shell, which was enclosed by two plasma shocks. A laser-plasma experiment can thus reproduce a configuration that is similar to that discussed in Refs. \cite{Vishniac94}, which gives rise to the nonlinear TSI.

Collisionless shocks are mediated by collective electromagnetic fields rather than by binary collisions between particles. The nature of the electromagnetic field distribution and the structure of the shock transition layer strongly depend on  plasma parameters such as the magnetization and the ratio between the shock speed and the light speed. A nonrelativistic plasma shock in an unmagnetized collisionless plasma is mediated by an electrostatic field \cite{Forslund71,Biskamp73}. The source of this electric field is the thermal diffusion of electrons from the dense downstream plasma into the upstream plasma, which leaves behind a positive net charge in the downstream region close to the shock. An electric field builds up that accelerates electrons from the upstream region into the downstream region. An equilibrium is established when the net outflow of electrons from the downstream region due to thermal diffusion is balanced by the electrons that are injected back into the downstream region by the electric field. The electric field is antiparallel to the density gradient. 

The electric field also affects the ions. It slows down the ions along the shock normal when they cross the shock and move into the downstream region. Such a structure resembles a shock in a collisional fluid and it is thus called an electrostatic shock. In contrast to fluid shocks, electrostatic shocks can reflect some of the incoming upstream ions and they do not fully thermalize the ions that cross the shock. Some of the downstream ions are accelerated upstream by the ambipolar electric field and such a structure is known as a double layer. A plasma shock is usually a combination of a double layer and of an electrostatic shock \cite{Hershkowitz81,Sorasio06}.  

\begin{figure}
\includegraphics[width=\columnwidth]{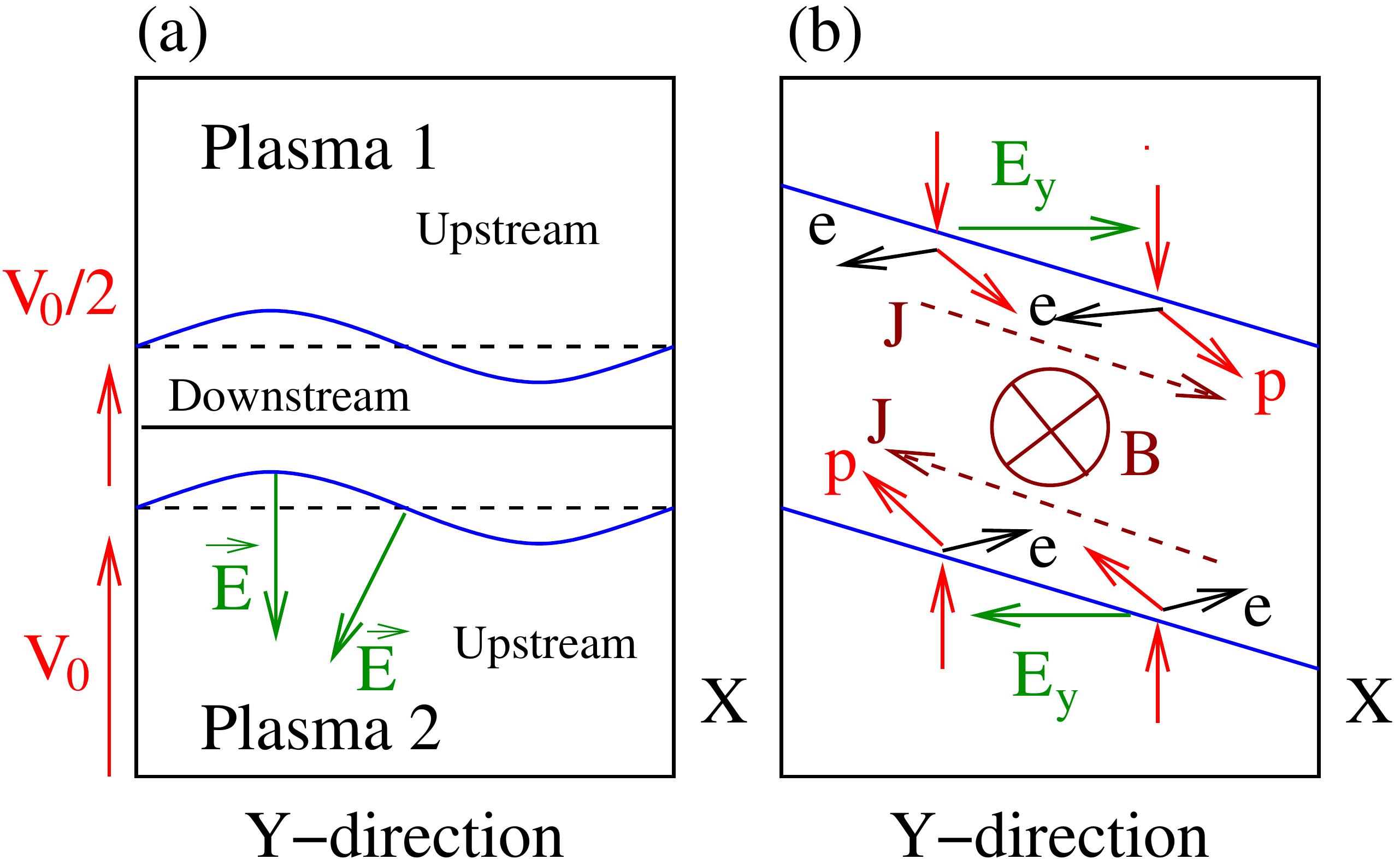}
\caption{(Color online) (a) The collision of the equally dense plasma 1 (at rest) and plasma 2 (speed $v_0$) yields two plasma shocks (blue sinusoidally varying curves), which enclose a plasma shell. Each shock is displaced along x with respect to its average plane (horizontal dashed black lines). The electric field $\mathbf{E}$, which mediates the plasma shock, points along the shock normal. Panel (b) focuses on the center of (a), where the horizontal electric field component $E_y$ is strongest. Ions (p) and electrons (e) are deflected into opposite directions as they cross a shock and the resulting net current $\mathbf{J}$ yields a magnetic $\mathbf{B}$-field.}\label{fig1}
\end{figure}

The question arises whether or not an instability, which is similar to the TSI, can develop in a collisionless laboratory plasma. Here we study this scenario by means of a particle-in-cell (PIC) simulation \cite{Dawson83} study with the EPOCH code \cite{Ridgers12}. Consider the case depicted in Fig. \ref{fig1}(a), where an upstream medium (plasma 2) is moving at a uniform flow speed towards the shock (lower blue sine curve). The shock boundary is displaced along this direction and the amplitude of the displacement varies sinusoidally with y. The same holds for the electric field vector $\mathbf{E}$ that mediates this plasma shock. 

The direction of $\mathbf{E}$ varies with y, while the ram pressure force of the upstream flow points along its uniform flow velocity vector; the latter is anti-parallel to the average shock normal. Hence the thermal pressure does not cancel out the ram pressure everywhere. The fluids are deflected into opposite horizontal directions by both shocks along a vertical cut in Fig. \ref{fig1}(a) and the deflection angle increases with an increasing obliquity of the shock boundary. Some important consequences of this deflection are revealed by our simulation study.

The plasma shocks are created in the simulation by the collision of two plasma clouds, each consisting of protons and electrons with the mass ratio $m_p / m_e = 1836$. The electrons and protons of each cloud have an equal density $n_0$ and mean speed modulus. Each cloud is thus initially charge- and current-neutral and we can set the electric $\mathbf{E}$-field and the magnetic $\mathbf{B}$-field to zero at the time $t=0$. 

Plasma 1 is at rest in the simulation frame and the mean speed of plasma 2 is $v_0$. The electron temperature $T_e = 1.74 \times 10^7$ K and the proton temperature $T_p=T_e/5$ are representative for laser-produced plasma. Initially, the plasma clouds are spatially separated and they are in contact at a boundary $x_B(y)$ with a sinusoidally varying displacement from a straight average boundary. Plasma 2 moves into plasma 1 for increasing times $t>0$ and a thin shell forms, which resembles that depicted in Fig. \ref{fig1}(a). The ions slow down as they cross the plasma shock and their piling up increases the density of the thin shell. The density increase depends on the Mach number of the cloud velocity with respect to the ion acoustic speed $c_s = {(\gamma_c k_B (T_e + T_p) / m_p)}^{0.5}$ ($k_B$: Boltzmann's constant) and measured in the reference frame of the shell. Typical values range from just over $2n_0$ to $3n_0$ for slow planar plasma shocks \cite{Forslund71,Dieckmann13}. The ion acoustic speed is $c_s \approx 5.4 \times 10^5$ m/s for our plasma parameters and the collision speed is $v_0 = 3.8c_s$. We have assumed that the adiabatic index is $\gamma_c = 5/3$ in order to obtain a reference value for $c_s$. The degrees of freedom in a plasma and hence the adiabatic constant depend on the field distribution. 

The proton Debye length $\lambda_{D} = {(\epsilon_0 k_B T_p / n_0 e^2)}^{1/2}$ normalizes space. The proton plasma frequency $\omega_{p}={(n_0 e^2/\epsilon_0 m_p)}^{1/2}$ normalizes time. The electric fields are normalized to $m_p v_{th} \omega_{p} / e$ and the magnetic fields to $m_p \omega_{p} / e$, where $v_{th}={(k_B T_p / m_p)}^{1/2}$ is the protons' thermal speed. The densities are expressed in units of $n_0$. 

The simulation resolves the interval $-864 \le x \le 864$ along the collision direction and $0 \le y \le 180$ perpendicular to it. The boundary condition at $x=-864$ is open, that at $x=864$ is reflecting and the boundary conditions along $y$ are periodic. The collision boundary is defined by $x_B(y) = 1.44 \cdot \sin{(2 \pi y / 90)}$ and it is thus centered around $x=0$. Plasma 1 initially occupies the interval $x_B(y) \le x \le 864$. Plasma 2 is located in the interval $-864 \le x < x_B(y)$. Momentum conservation implies that the shell will propagate to increasing values of $x$ with the speed $\approx v_0 / 2$. The simulation grid resolves x by 2400 cells and y by 250 cells. The side length of each cell is thus $0.72 \lambda_{D}$. Electrons and protons are each represented by 200 particles per cell at the simulation's start and we do not inject new particles while the simulation is running. The total simulation time is $t_{sim}=68$. 

\begin{figure}[ht]{}
\includegraphics[width=\columnwidth]{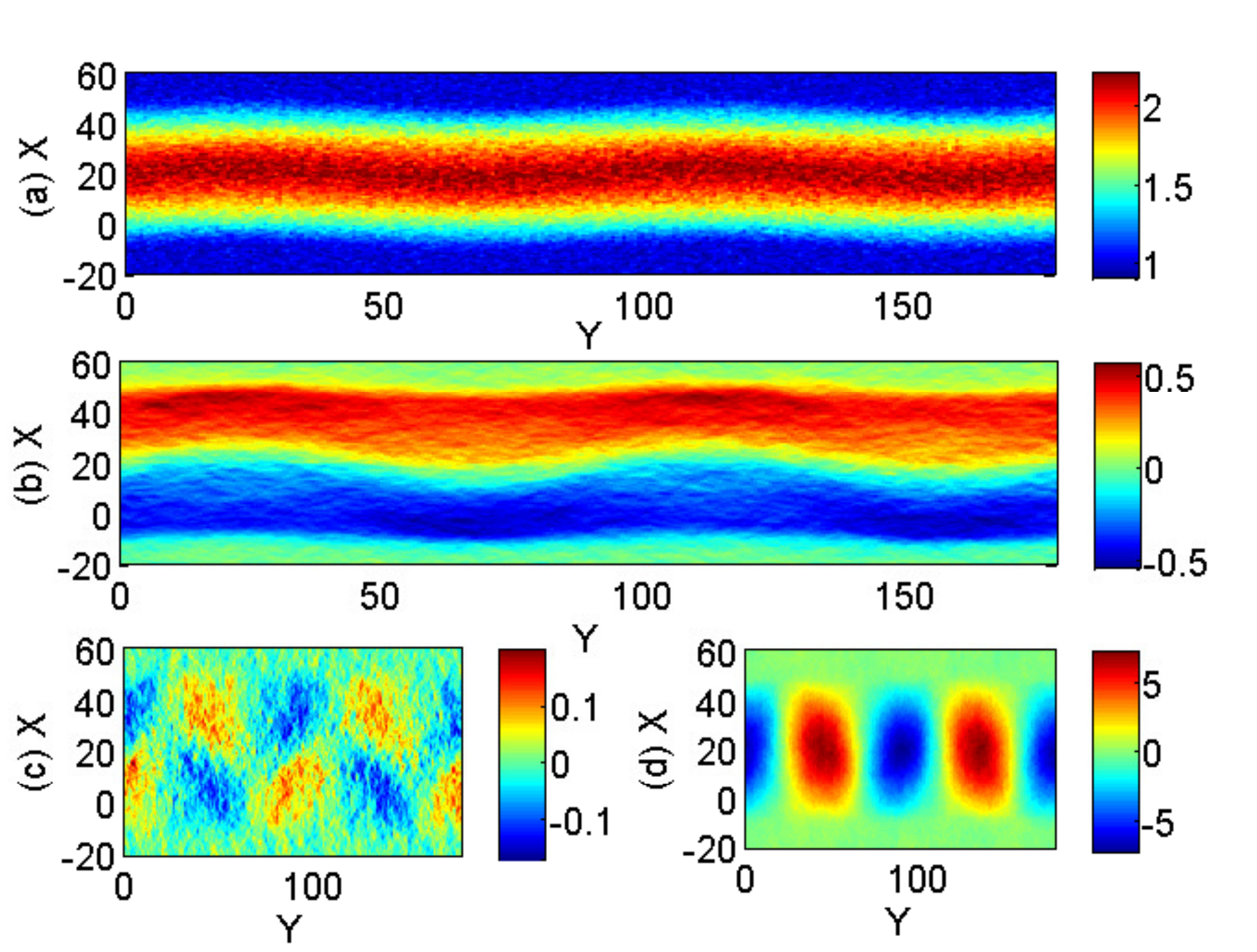}
\caption{(Color online) The thin shell at the time $t=4.5$: Panel (a) show $n_p(x,y)$. Panels (b,c,d) show the $E_x$, $E_y$ and $B_z$ components, respectively. The value of $B_z$ has been multiplied by the factor $10^5$.}\label{fig2}
\end{figure}

Figure \ref{fig2} displays the total proton density distribution $n_p(x,y)$ together with $E_x(x,y), E_y(x,y)$ and $B_z(x,y)$ at the time $t=4.5$. Movie 1 \cite{Mov1} shows the time-evolution of $n_p(x,y)$ throughout the simulation and Movie 2 shows that of $B_z(x,t)$ \cite{Mov2}.
The shell is characterized in Fig. \ref{fig2}(a) by a peak proton density $\sim 2.3$, which implies that the inflowing ions have been slowed down and compressed by the positive potential of the shell. The displacement along x of the shell is comparable to the initial amplitude of $x_B(y)$ at this time. The in-plane electric field vector $(E_x,E_y)$ varies with $y$ and the peak amplitude of $E_y$ is about one half of that of $E_x$. 

The inflowing electrons and protons are deflected into opposite directions when they cross a plasma shock with its strong associated $E_y$-field. The therefrom resulting net current in the simulation plane generates the observed $B_z \neq 0$ within the thin shell. We can see this from Fig. \ref{fig1}(b), which corresponds to a position $y=45$ in Fig. \ref{fig2}. The current, which is generated by the horizontal deflection of the electrons and protons, points to the left at the lower plasma shock and to the right at the upper plasma shock at $y=45$. Such a current distribution yields a magnetic field, which points into the plot's plane and, thus, into the same direction as the positive z-axis in Fig. \ref{fig2}(d). The periodicity of the distributions of $E_y$ and $B_z$ equals that of the initial boundary oscillation $x_B(y)$ and their phases are shifted by $\pi/2$, which is further evidence for a magnetic field generation by the particle deflection by $E_y$. 

\begin{figure}
\includegraphics[width=\columnwidth]{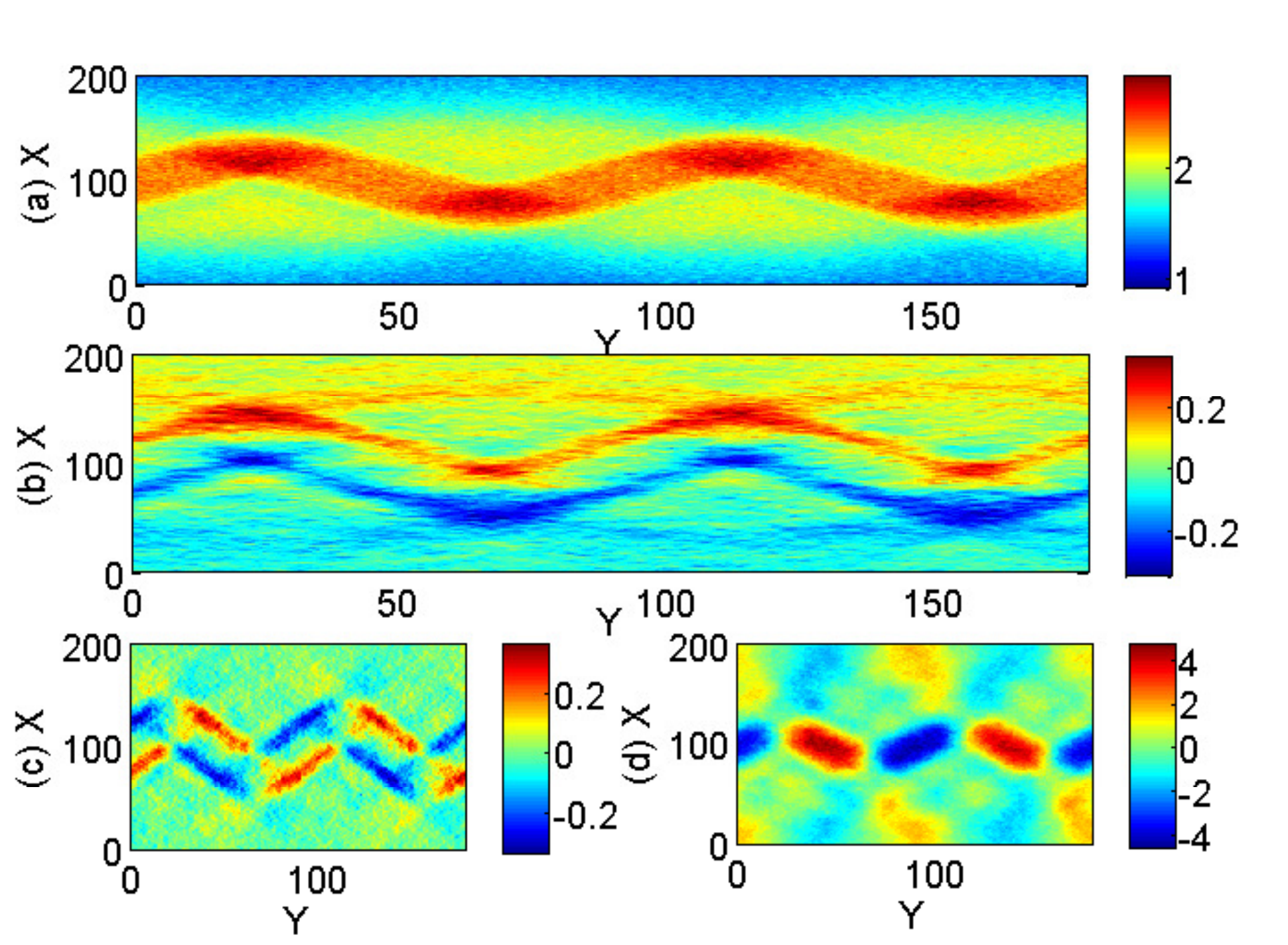}
\caption{(Color online) The thin shell at the time $t=24$: Panel (a) show $n_p(x,y)$. Panels (b,c,d) show the $E_x$, $E_y$ and $B_z$ components, respectively. The value of $B_z$ has been multiplied by $10^5$.}\label{fig3}
\end{figure}

Figure \ref{fig3} shows the simulation results at $t_{sim}=24$. 
The shell has expanded along $x$ and shows a sinusoidal displacement with the same phase and period as that of $x_B(y)$. The amplitude of the oscillations of the thin shell along $x$ is about 20, which exceeds that of $x_B(y)$ by a factor 15. The proton density peaks at extrema of the oscillations of the thin shell. The strongest structures in the electric field in Fig. \ref{fig3}(b) outline the proton density jump between the thin shell and the surrounding plasma and they demarkate the locations of both plasma shocks. The $E_y$ component in Fig. \ref{fig3}(c) shows piecewise linear patches that are correlated with the locations of the plasma shocks. The peak amplitudes of $E_x$ and $E_y$ are comparable in strength. The extent along y of the patches in Fig. \ref{fig3}(c) is about 30 and their position along x varies by about the same value. The angle between the electric field $(E_x,E_y)$ and the average normal of each shock is thus about $\pi / 2$ within these intervals. The magnetic field structures in Fig. \ref{fig3}(d) are strongest within the thin shell and they remain closely correlated with the electric $E_y$ field patches. The numerical values of $B_z$ correspond to the ratio between the proton cyclotron frequency and the proton plasma frequency and the peak value is about $5\times 10^{-5}$. The effects of the magnetic field on the protons is negligible.

\begin{figure}
\includegraphics[width=\columnwidth]{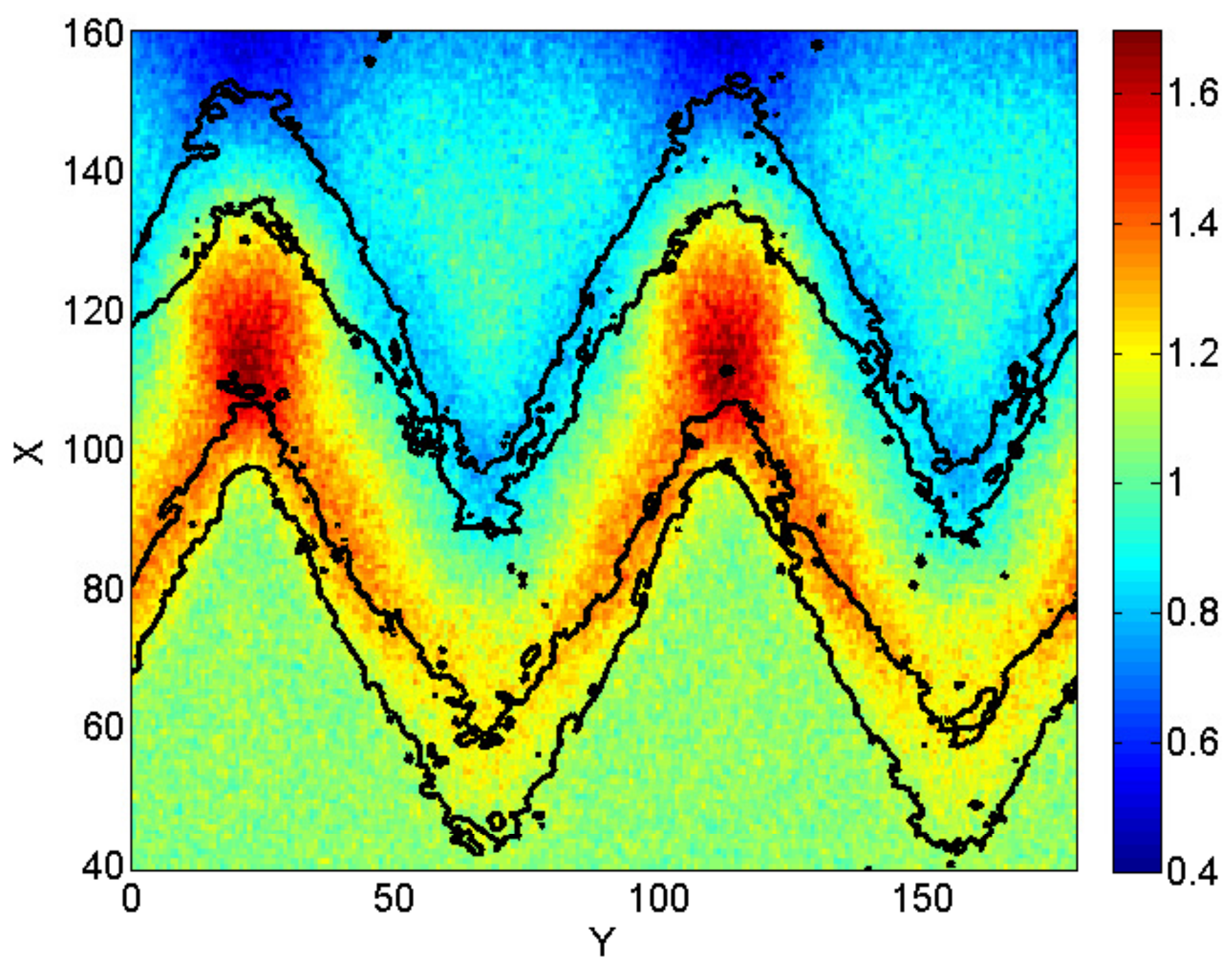}
\caption{(Color online) The density distribution of the protons of plasma 2 at the time $t=24$. Overplotted is the contour $0.23$ of the electric field modulus ${(E_x^2 + E_y^2)}^{1/2}$.}\label{fig4}
\end{figure}

Figure \ref{fig4} compares the electric field contour ${(E_x^2 + E_y^2)}^{1/2} = 0.23$ with the proton density distribution of plasma 2. These protons flow towards increasing values of $x$. 
The proton density at low values of $x$ is close to 1. These protons correspond to the upstream plasma, before it has encountered the lower plasma shock. The lower shock is outlined by the lower pair of countour lines. The proton density hardly increases when the protons cross the convex parts of the lower shock at $y\approx 65$ or $y\approx 155$. The density is increased substantially when the protons cross the concave parts of the lower shock at $y \approx 25$ or $y\approx 110$. The density changes gradually as a function of $y$ in the intervals where the lower shock has a constant tilt angle relative to the incoming upstream plasma flow direction. The density change is symmetric with respect to the locations of the extrema of the shell's oscillation. The density change is also closely correlated with the electric field contours. The proton distribution evidences that the inflowing protons are deflected by the electric $E_y$-component towards those y-intervals, where the lower shock is concave. The protons accumulate at this location and their density increases to a value $\approx 1.7$. This proton flow direction matches that in Fig. \ref{fig1}(b). 

We find smaller peaks of the proton density above the upper plasma shock, which is outlined by the upper pair of electric field contour lines. The protons, which have crossed the thin shell and have reached its opposite side, are reaccelerated by the upper shock and escape into the upper upstream region. This reacceleration is accomplished by the double layer component of the upper plasma shock. The tilted electric field channels these re-accelerated protons into the areas centered at $y\approx 65$ and $x\approx 120$ as well as  $y\approx 155$ and $x\approx 120$.

The steady deflection of inflowing protons by the horizontal electric field component towards the concave shocks will increase further their density at these locations; the proton distribution in the Figs. \ref{fig3} and \ref{fig4} is not in a steady state (See also movie 1). However, the proton density difference between the position $(x,y)=(20,115)$ and $(x,y)=(45,95)$ in Fig. \ref{fig3}(a) is already about 0.5 and, thus, comparable to the difference between the upstream plasma and the downstream plasma. The ambipolar electric field, which is associated with the proton density variations along the thin shell, will eventually become strong enough to accelerate protons from the dense parts of the thin shell to the dilute ones. 

\begin{figure}
\includegraphics[width=\columnwidth]{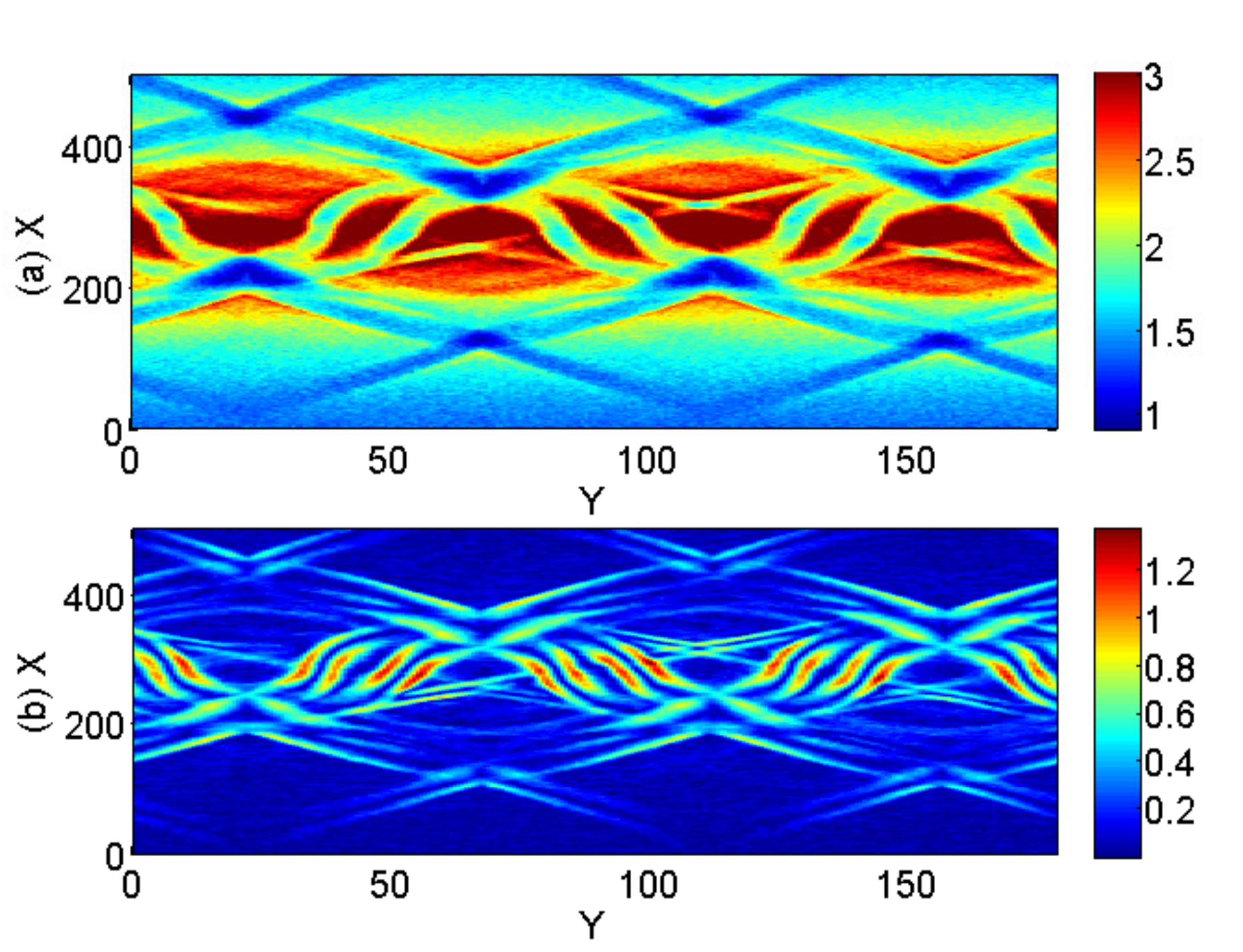}
\caption{(Color online) The thin shell at the time $t=68$: Panel (a) shows $n_p(x,y)$. Panel (b) shows ${(E_x^2 + E_y^2)}^{1/2}$.}\label{fig5}
\end{figure}

The impact of this secondary ambipolar electric field on the protons in the thin shell is evidenced by Fig. \ref{fig5}, which corresponds to the simulation time $t=68$.
Figure \ref{fig5}(a) reveals a complicated proton density distribution within the thin shell. Most protons are still located behind the concave shocks at $(x,y)=(20,250)$, $(x,y)=(115,250)$, $(x,y)=(70,350)$ and $(x,y)=(160,350)$. New proton density maxima have appeared, which are centered at the positions $x\approx 300$ and $y$ = 0, 45, 90 and 135. These $y$-coordinates correspond to those locations where the initial contact boundary $x_B(y)=0$ and where the density of the thin shell was lowest in Fig. \ref{fig3}(a). The density decreases from 3 to about 1.8 near these maxima. These density jumps give rise to the strongest electric fields in Fig. \ref{fig5}(b). The localized proton density depletions inside of the thin shell and their associated electric fields appear to be stable and they convect with the thin shell. This is evidenced by Movie 3 \cite{Mov3}, which animates in time ${(E_x^2 + E_y^2)}^{1/2}$. 

\begin{figure}
\includegraphics[width=\columnwidth]{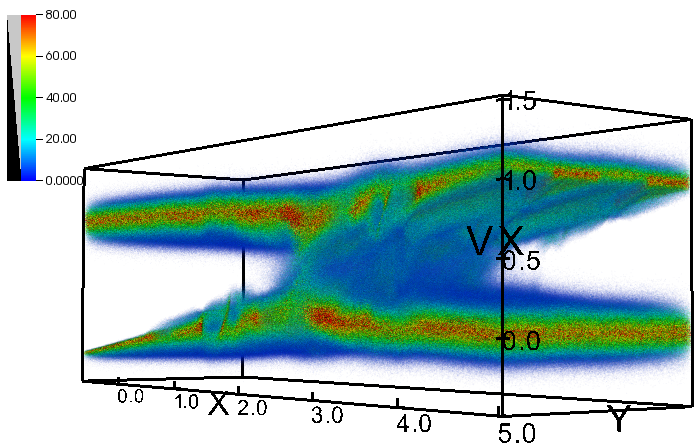}
\caption{(Color online) The proton phase space density distribution $f_i(x,y,p_x)$ at the time $t=68$: The color range is linear. The x-position is given in units of $100\lambda_D$ and $v_x$ is given in units of $v_0$. The full width of the y-axis is displayed.}\label{fig6}
\end{figure}

Figure \ref{fig6} demonstrates that two electrostatic shocks, which separate the downstream region at $x\approx 300$ from the upstream regions at large and low values of $x$, are forming at the time $t=68$.
The bulk of the protons is still propagating with the interpenetrating ion clouds with $v_x \approx 0$ or $v_x \approx v_0$. A significant population is, however, also filling up the phase space interval with $x\approx 300$ and $v_x \approx v_0 / 2$. The ion phase space structures at $-100 < x < 200$ and $v_x\approx 0$ and $400 < x < 500$ and $v_x \approx v_0$ are fed by the shock-reflected ions and by ions that crossed the downstream region and were re-accelerated by the second shock.

Summary: We have tested with a two-dimensional particle-in-cell simulation the stability of a thin shell of collisionless plasma, which is enclosed by plasma shocks. The thin shell was displaced along the normal of its average plane. The displacement varied sinusoidally as a function of the position along the thin thell. We have demonstrated that the thin shell is unstable. The amplitude of the displacement grew 15-fold. The thin shell followed a piecewise linear spatial distribution when the thin-shell instability saturated. The continuing accumulation of protons at the density maxima led to the formation of large-amplitude density modulations within the thin shell. These proton density depletions, which are sustained by strong electric fields, will probably become phase space structures known as proton phase space holes \cite{Luque05,Eliasson06}. The deflection of the upstream electrons and protons, which crossed the plasma shock, into opposite directions resulted in a net current within the thin shell that generated a magnetic field.

\textbf{Acknowledgements:} MED acknowledges financial support from CRAL (ENS). RW und DF acknowledge support by the French national program of high energy (PNHE). This work was supported by grant ENE2013-45661-C2-1-P from the Ministerio de Educaci\'on y Ciencia, Spain, and grant PEII-2014-008-P from the Junta de Comunidades de Castilla-La Mancha. This research was conducted using the resources of High Performance Computing Center North (HPC2N).

\end{document}